\documentclass[graybox]{svmult}

\usepackage{mathptmx}       % selects Times Roman as basic font
\usepackage{helvet}         % selects Helvetica as sans-serif font
\usepackage{courier}        % selects Courier as typewriter font
\usepackage{type1cm}        % activate if the above 3 fonts are
\usepackage{makeidx}         % allows index generation
\usepackage{graphicx}        % standard LaTeX graphics tool
\usepackage{multicol}        % used for the two-column index
\usepackage[bottom]{footmisc}% places footnotes at page bottom

\makeindex             % used for the subject index
\begin{document}

\title{Jan Tinbergen's legacy for economic networks: from the gravity model to quantum statistics}
\titlerunning{Jan Tinbergen's legacy for economic networks} %for an abbreviated version of
\author{Tiziano Squartini and Diego Garlaschelli}
\institute{Tiziano Squartini \at Instituut-Lorentz for Theoretical Physics, Leiden Institute of Physics, University of Leiden, Niels Bohrweg 2, 2333 CA Leiden, \email{squartini@lorentz.leidenuniv.nl}
\and Diego Garlaschelli \at Instituut-Lorentz for Theoretical Physics, Leiden Institute of Physics, University of Leiden, Niels Bohrweg 2, 2333 CA Leiden, \email{garlaschelli@lorentz.leidenuniv.nl}}
\maketitle

\abstract{Jan Tinbergen, the first recipient of the Nobel Memorial Prize in Economics in 1969, obtained his PhD in physics at the University of Leiden under the supervision of Paul Ehrenfest in 1929. 
Among many achievements as an economist after his training as a physicist, Tinbergen proposed the so-called Gravity Model of international trade. 
The model predicts that the intensity of trade between two countries is described by a formula similar to Newton's law of gravitation, where mass is replaced by Gross Domestic Product. 
Since Tinbergen's proposal, the Gravity Model has become the standard model of non-zero trade flows in macroeconomics. 
However, its intrinsic limitation is the prediction of a completely connected network, which fails to explain the observed intricate topology of international trade. 
Recent network models overcome this limitation by describing the real network as a member of a maximum-entropy statistical ensemble. 
The resulting expressions are formally analogous to quantum statistics: the international trade network is found to closely follow the Fermi-Dirac statistics in its purely binary topology, and the recently proposed mixed Bose-Fermi statistics in its full (binary plus weighted) structure. 
This seemingly esoteric result is actually a simple effect of the heterogeneity of world countries, that imposes strong structural constraints on the network.
Our discussion highlights similarities and differences between macroeconomics and statistical-physics approaches to economic networks.}

\section{Introduction}
\label{sec:1}
Over the last fifteen years, there has been an ever-increasing interest in the study of networks across many scientific disciplines, from physics to biology and the social sciences \cite{networks}. 
Economics is no exception. 
Empirical \cite{yoshi,myreview} and theoretical \cite{jackson,goyal} analyses of economic networks have been growing steadily, gradually encompassing different scales: from `microscopic' networks of individual agents and financial assets \cite{jackson,tiziana}, through `mesoscopic' networks of firms, banks and institutions \cite{yoshi,myinvestments,myearly}, to `macroscopic' networks of world countries and economic sectors \cite{mywtw,productspace,pietronero}. 
An unprecedented element of continuity across these different economic scales has been the search for empirical laws characterizing real-world networks and the subsequent introduction of simple models aimed at reproducing the observed `stylized facts'. 
Placing observations, rather than theoretical postulates, at the starting point of scientific investigations is probably the main positive outcome of the interaction between economists and  physicists, an interaction that - over the last two decades - has given rise to the controversial field of `Econophysics'. 
The interdisciplinary study of economic networks is another very fruitful result of this interaction. 
The added value of using the network approach to economic problems is the possibility to investigate indirect effects arising as the combination of many pairwise interactions between economic agents or units.
The prototypical example is the study of \emph{systemic risk}, i.e. the risk of a system-wide cascade of defaults of banks or institutions connected to each other in a financial network, as opposed to traditional measures of risk for single financial entities.

Despite the `network approach' is relatively recent, much earlier studies in Economics already recognized the importance of (what we now call) socio-economic networks, even if this knowledge was more or less dispersed across sub-fields that used to be largely disjoint.
An important example is the so-called Gravity Model \cite{anderson}. 
The name originates from a loose analogy with Newton's law of gravitation, which states that the gravitational force between two objects is proportional to the product of their masses, and inversely proportional to the square of the distance between them.
Strictly applying the analogy to the economic setting, the Gravity Model  (see \cite{anderson} for an excellent review) assumes that a `mass' $S_i$ of goods (or services, or factors of production such as labor) supplied at an origin $i$ is `attracted' to a mass $D_j$ of demand for such goods located at a destination $j$. This attraction generates a flow $F_{ij}$ of goods, but the flow is reduced by the geographic distance $d_{ij}$ between origin and destination as follows
\begin{equation}
F_{ij}= K\frac{S_i\:D_j}{d^2_{ij}}.
\label{formula1}
\end{equation}
where $K$ is a global free parameter to be fitted to real data.
The Gravity Model predicts larger fluxes between closer and `bigger' (in terms of the size of supply and demand) locations, exactly in the same way as the gravitational force is stronger between closer and more massive objects.

The use of the Gravity Model was pioneered by Ravenstein \cite{ravenstein} in studies of migration patterns, where flows represent movements of people, and $S_i$ and $D_j$ are mainly determined by the sizes of the two populations located at the origin and destination. 
Jan Tinbergen, the first recipient of the Nobel Memorial Prize in economics, was instead the first to use the Gravity Model to explain international trade flows \cite{jantin}.
In this case, flows represent movements of goods among world countries, and $S_i$ and $D_j$ are expected to be determined by the values of the Gross Domestic Product (GDP) of the countries of origin and destination, i.e. $S_i=D_i=GDP_i$.

Indeed, it is precisely as the first data-driven model of trade that the Gravity Model acquired its great popularity, presumably because of the accuracy with which it predicts observed trade fluxes.
On the other hand, international trade is also one of the best examples of economic systems that have been intensively studied, using a completely different `network' approach, over the last decade.
This coincidence makes the International Trade Network a useful example to compare traditional (economic) and recent (network) approaches when applied to the same system, which in this case is also an empirically well documented one. 
For this reason, in what follows we will focus on Jan Tinbergen's Gravity Model of trade, its successes and limitations, and the more recent approaches that overcome some of these limitations.
As another curious coincidence, these very different frameworks have a common element: the modeling of an economic network in close analogy with physical laws, from gravitation to statistical physics and quantum statistics.

\section{Jan Tinbergen and the Gravity Model of trade}\label{sec:gravity}
In a slightly (and not fully) generalized form, the Gravity Model of international trade states that the expected amount of trade from country $i$ to country $j$ is 
\begin{equation}
\langle w_{ij}\rangle= K\frac{GDP_i^{\alpha}\:GDP_j^{\beta}}{d_{ij}^\gamma}
\label{formula2}
\end{equation}
where $d_{ij}$ is the geographic distance between countries, and $\alpha$, $\beta$ and $\gamma$ are additional (besides $K$) free parameters. 
The angular brackets in eq.(\ref{formula2}) denote an expected value: this means that the model is not intended to be a deterministic one, since real data will obviously deviate from the postulated expression. So, strictly speaking, the model predicts that the realized amount of trade is $w_{ij}=\langle w_{ij}\rangle+\epsilon$ where $\epsilon$ is an error term with zero mean (if a linear regression to the observed trade flows is used), or alternatively $w_{ij}=\langle w_{ij}\rangle\cdot\eta$ where $\eta$ is an error term with unit mean (if a linear regression to the \emph{logarithm} of the observed trade flows is used).
In both cases, the fitted values of the parameters are usually around $\alpha\approx\beta\approx\gamma\approx 1$ \cite{anderson}.
Further extensions of eq.(\ref{formula2}) include additional factors either favouring or suppressing trade.\footnote{Examples of favouring factors are: trade agreements, membership to common economic groups, shared geographic borders, common currency, etc. Examples of suppressing factors are: embargoes, trade restrictions, and other factors representing a trade friction.\label{foot:regressors}}
Despite the inclusion of these additional factors improves the fit, the main factor determining trade flows remains the GDP, followed by geographic distances. 
So eq.(\ref{formula2}), with exponents $\alpha\approx\beta\approx\gamma\approx 1$, captures the basic lesson learnt from real trade data and makes the Gravity Model closer to the expression for the gravitational energy ($\gamma=1$) than to the one for the gravitational force ($\gamma=2$).

Obviously, there is absolutely nothing fundamental in the formal analogy between the empirical laws of trade (or any other economic flux) and gravity, and no profound reason why these laws should bear any mathematical similarity at all. 
Rather, the deep similarities must be looked for at different levels:

\begin{itemize}
\item A first analogy involves the implicit use of \emph{symmetry} in both cases: both eq. (\ref{formula2}) and Newton's law, state that, \emph{all else being equal}, only mass/GDP and distance determine the amplitude of the interaction. In physics, this means that Newton's law holds \emph{in vacuum}, i.e. in absence of anything else that can interact gravitationally with the two objects. In macroeconomics, this means that eq.(\ref{formula2}) holds in absence of other factors affecting trade, such as the additional regressors we mentioned in footnote \ref{foot:regressors}.\\
\item A second similarity is the \emph{qualitative dependence} on the key quantities: both laws assume that interaction amplitudes increase with increasing mass/GDP, and decrease with increasing distance. In principle, there is an infinity of quantitative ways (functional forms) to implement this qualitative idea. Accidentally, the functional forms describing gravitation and trade turn out to be very similar, but their qualitative analogy would have held even if the two mathematical expressions were different. In some sense, this makes the qualitative analogy more fundamental than the mathematical one.\\
\item The above consideration leads us to a third analogy, i.e. the \emph{phenomenological} character common to eq. (\ref{formula2}) and Newton's formula. 
In both laws, the particular functional form that implements the previous theoretical arguments is established on the basis of its success in reproducing real data, and thus \emph{a posteriori}.
Other functional forms, while possible \emph{a priori} on the basis of the above two points, must be discarded if they do not explain observations.
Only after they were widely accepted as powerful empirical laws explaining observations, Netwon's law and the Gravity Model became the `target' of more general and abstract theories. For instance, to be acceptable, Einstein's theory of General Relativity \emph{must} reduce to Netwon's law in the appropriate circumstances, and micro-founded economic models must generate the Gravity Model when aggregated at the macro level \cite{anderson}.
\end{itemize}
In our view, the above epistemological analogies are even more fundamental than the (accidental) mathematical analogy between eq.(\ref{formula2}) and Netwon's law.
Another deep connection between physics and economics exists at a personal level: Jan Tinbergen, the founder of the Gravity Model of trade, was a physicist before becoming an economist.

\begin{figure}[t!]
\sidecaption
\includegraphics[scale=.36]{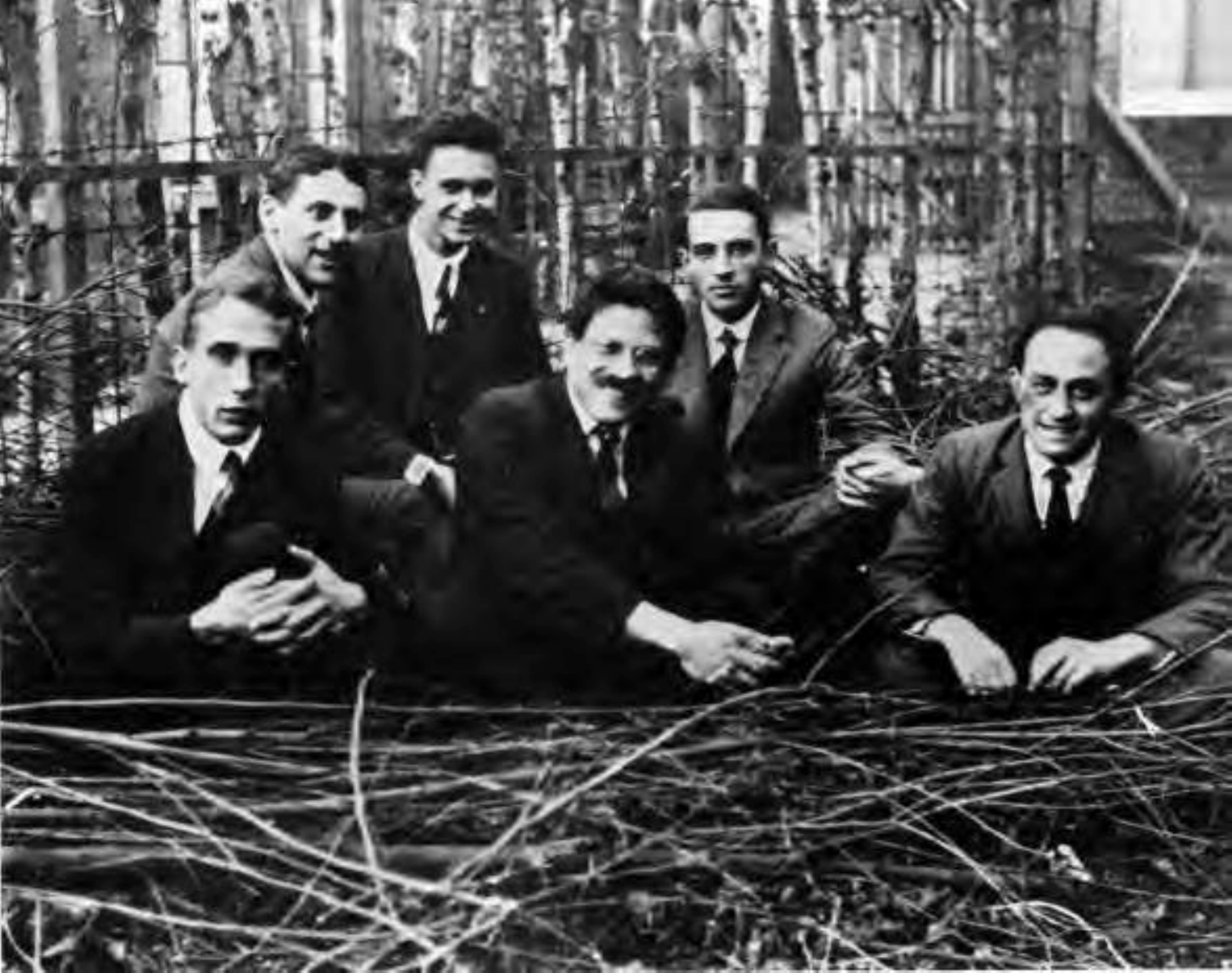}
\caption{Ehrenfest's students, Leiden 1924. Left to right: Gerhard Heinrich Dieke, Samuel Abraham Goudsmit, Jan Tinbergen, Paul Ehrenfest, Ralph Kronig, and Enrico Fermi (copyright $\copyright$ Chicago University Press).}
\label{fig:0}
\end{figure}

Born in Den Haag, the Netherlands, in 1903, Jan Tinbergen started his studies in mathematics and the natural sciences at the University of Leiden, soon after graduating from high school in 1921 with the highest honors. 
In Leiden, he later started a PhD in physics under the supervision of Paul Ehrenfest, who was then professor in Theoretical Physics (see fig.\ref{fig:0}). Tinbergen became Ehrenfest's assistant, the private tutor of Ehrenfest's son, and a frequent visitor of Ehrenfest's house, that was regularly visited also by Einstein, Bohr, Heisenberg, Fermi and Pauli. 
Tinbergen had always been attracted by economics, and Ehrenfest was interested in the analogies between economics and physics. This resulted in Tinbergen's PhD thesis, entitled `Minimum Problems in Physics and Economics' and defended in 1929.
Shortly after, despite Ehrenfest had repeatedly tried to convince him to remain a physicist, Tinbergen started a brilliant career as an economist.
His pioneering views led him to introduce Econometrics, a synthesis between mathematics, economic theory and statistics. In Tinbergen's view, economic theory should formulate hypotheses translated into mathematical relations that are then statistically tested on empirical data. 
This distinctive quantitative approach was almost surely due to Tinbergen's graduate training as a physicist. 
His idea of introducing a quantitative model of international trade flows is clearly in line with this approach.
Jan Tinbergen's career culminated in 1969 when he received the first \emph{Bank of Sweden Prize in Economic Sciences in Memory of Alfred Nobel} or shortly \emph{Nobel Memorial Prize in Economics}, often mistakenly referred to as the `Nobel Prize in Economics' (that, strictly speaking, does not exist). 

As a physicist, Jan Tinbergen of course knew Netwon's law very well, a knowledge that might have facilitated making a mathematical connection to the study of international trade. 
But, we believe, his idea of using the Gravity Model in his research as an economist was most probably triggered by the deeper similarities, discussed above, at the epistemological level.
Jan Tinbergen's familiarity with the scientific method universally used in physics is probably the reason why his many achievements as an economist are all characterized by a strong quantitative approach and a clear focus on empirical data.
Without trying to distort scientific and personal history, we might therefore presume that Tinbergen's view was not too far from what, in modern jargon, are the inspiring concepts of `Econophysics'.
His Gravity Model of trade can also be regarded as the first model of the system that, in the more recent Complex Networks literature, has been intensively studied under the names of `International Trade Network' (ITN) or `World Trade Web' (WTW) \cite{mywtw,serrano,mypre1,mypre2,myjeic,mytriadic}.
Therefore, in our view, Jan Tinbergen's pioneering work deserves full attention from the scientific community active in Econophysics and Network Science.
Along these lines, the Gravity Model and the ITN, both still under intense investigation, can be considered two brilliant examples of how ideas from physics can fruitfully interact with economic problems.

\section{The network approach}\label{sec:network}
At the time of Tinbergen's analyses, data about international trade flows were of course much less accurate than today.
Missing data were the rule rather than the exception, so it was practically impossible to distinguish between the absence of data documenting an existing trade relationship and a `true' absence of the relationship itself in the real world.
While this confusion still cannot be completely eliminated, in modern databases \cite{gle,UNCOM} it only affects a few percent of the data.
A simple analysis of such databases yields to a systematic result: in each yearly snapshot of the ITN from the 50's until now, only 50-60\% of the total pairs of world countries are found to be connected by trade relationship \cite{mywtw,mypre1}. 
With little error, pairs of countries that do not trade at all are the remaining fraction.

If we look again at eq. (\ref{formula2}), we immediately see that the observation of a `half-connected world' is in contrast with the predictions of the Gravity Model, as it cannot predict zero trade flows.\footnote{Strictly speaking, the introduction of an error term into eq.(\ref{formula2}) allows to have zero or even negative values. However, after fitting the model to the data, or simply in order to avoid the generation of precisely those unrealistic negative values, the variance of the error term is so small that zero trade flows have a vanishing probability.} 
Exactly as the gravitational force between any two masses (no matter how small or distant) is never zero according to Newton's law, the Gravity Model predicts that trade exchanges between any two countries (no matter how poor or distant) are always positive. 
However, while any two massive objects are indeed found to be attracted over cosmological distances in our Universe, the observation of an economically half-connected world implies that \emph{the Gravity Model fails in reproducing the missing links of the world trade network}.\footnote{In principle, also this limitation can be overcome if the Gravity Model is extended into the so-called \emph{zero-inflated} models \cite{zeroinflated,gravity_giorgio} that use eq.(\ref{formula2}) (or its generalizations) in a two-step procedure: first in order to estimate the probability of a trade connection, and then in order to estimate the intensity of the connection. However, recent analyses \cite{gravity_giorgio} have shown that this procedure provides a bad fit to the observed network: when used in order to estimate link probabilities and link intensities simultaneously, the Gravity Model turns out to be a very bad model.} 
In other words, if the set of existing connections (i.e. the \emph{topology}) of the ITN is preliminarly specified, then the Gravity Model succeeds in reproducing the magnitude of trade connections. But in general, it fails in reproducing the observed topology of the network. 

\begin{figure}[t!]
%\sidecaption
\includegraphics[scale=.4]{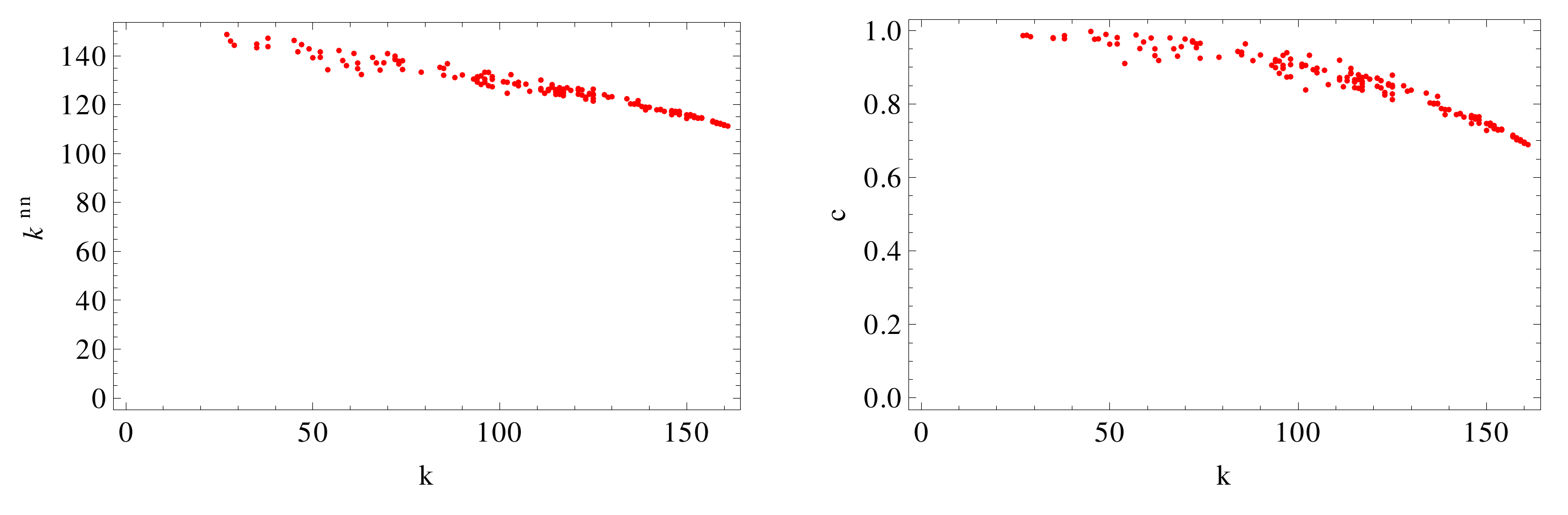}
\caption{The heterogeneous topology of the binary World Trade Web.
Left: average nearest neighbour degree $k^{nn}$ as a function of the degree $k$, for all vertices. Right: clustering coefficient $c$ as a function of the degree $k$. 
Data: UNCOMTRADE database \cite{UNCOM}, year 2001 ($N=162$ countries).}
\label{fig:binary}
\end{figure}

It is interesting to notice that the awareness of the importance of the \emph{binary} topology of a network, besides that of  \emph{weighted} structural properties, is a recent conquest of Network Science -- clearly absent at the time of Tinbergen. 
Motivated by this awareness, the analyses of the ITN carried out over the last decade have documented an intricate and heterogeneous topology. 
Let us for instance consider the \emph{undirected} version of the ITN, where two countries (the nodes, or \emph{vertices}, of the network) are connected by a \emph{link} (or \emph{edge}) if there exists at least a trade relationship (in any direction) between them.
In this network, the number of connections (the so-called \emph{degree}, denoted by $k$) of world countries is found to be very broadly distributed, with poor countries having only one or two connections (typically including the USA) and rich countries being connected to a significant number of partners, up to the total number of countries in the world.
This result is very robust, since the degree is found to systematically increase with the GDP \cite{mywtw}.
Moreover, (anti)correlations between the degrees of two trading partners are significant: the average degree of trade partners (the \emph{average nearest neighbour degree}, denoted by $k^{nn}$) is smaller for countries with larger degree \cite{mywtw,serrano} (see fig. \ref{fig:binary}).
This means that more connected (richer) countries trade with countries having on average a smaller number of partners, and less connected (poorer) countries trade with countries having on average a larger number of partners.
A similar result holds for the so-called \emph{clustering coefficient} (denoted by $c$) of a country, defined as the realized fraction of links (local link density) among the partners of that country.
Just like $k^{nn}$, $c$ is found to decrease as $k$ increases, meaning that more connected countries have a less interconnected neighbourhood, and less connected countries have a more interconnected neighbourhood \cite{mywtw,serrano} (see fig. \ref{fig:binary}).
All these topological properties can be generalized to the \emph{directed} version of the network (where links follow the direction of, say, exports), and similar results are found \cite{mypre1,myjeic}.

\begin{figure}[t!]
%\sidecaption
\includegraphics[scale=.43]{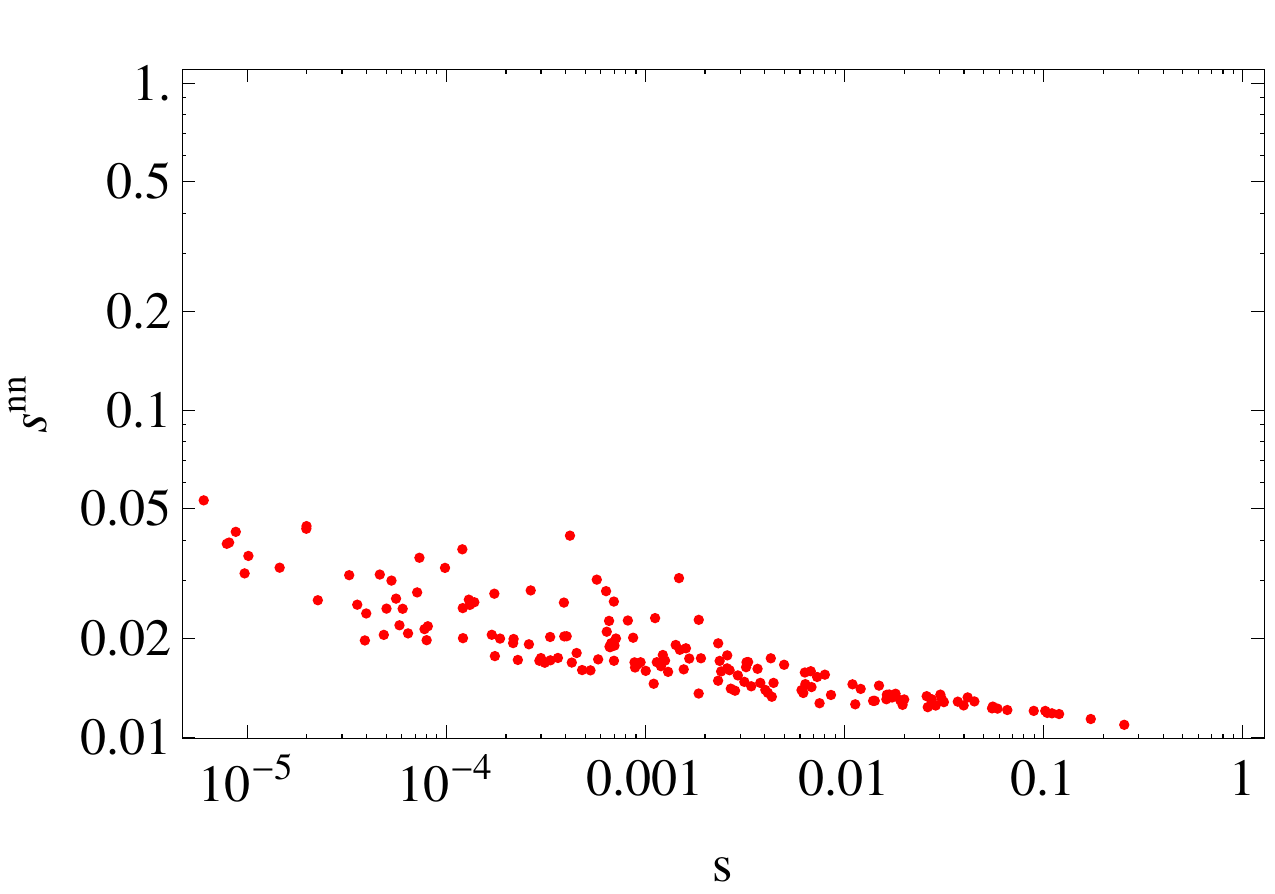}\includegraphics[scale=.44]{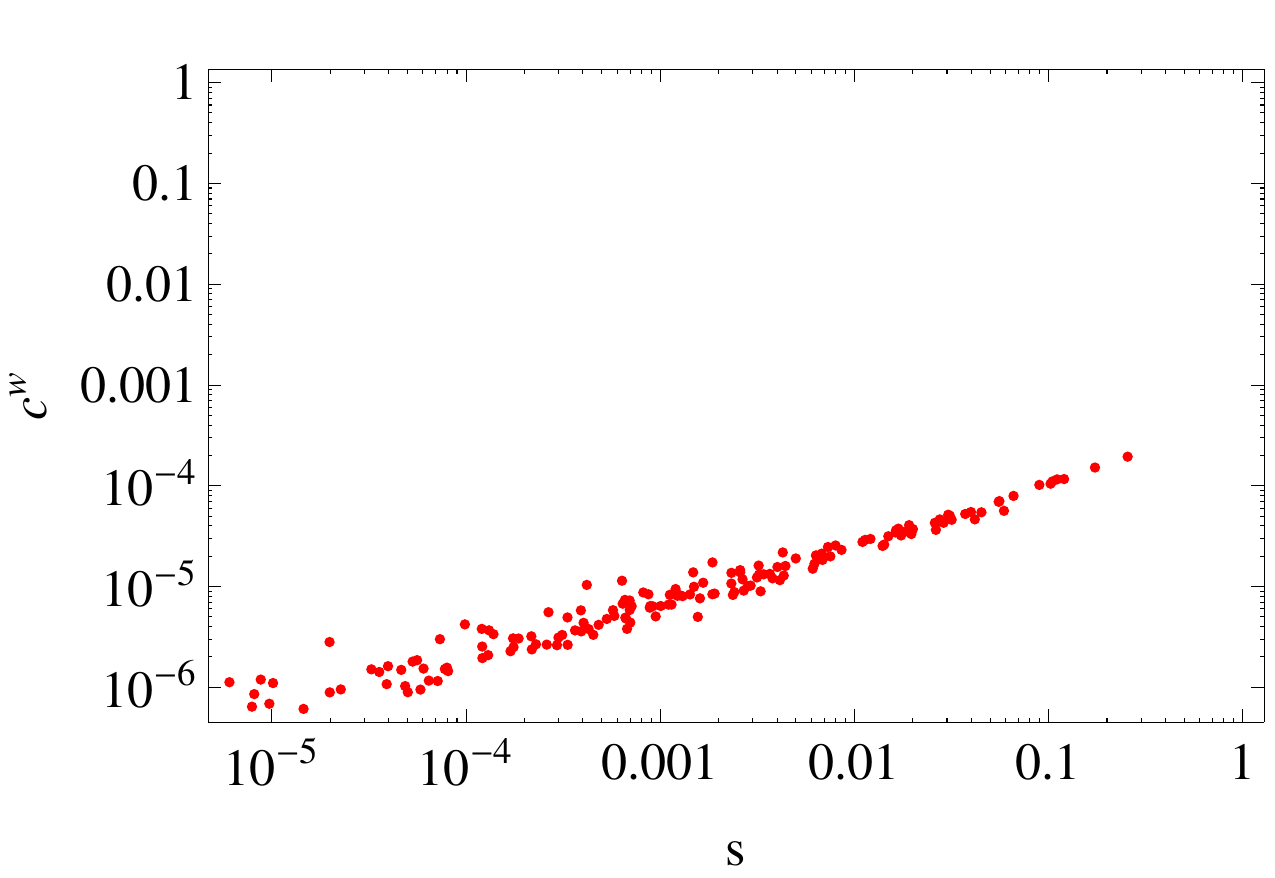}
\caption{The nontrivial structure of the weighted World Trade Web.
Left: average nearest neighbour strength $s^{nn}$ as a function of the strength $s$, for all vertices. Right: weighted clustering coefficient $c^w$ as a function of the strength $s$. 
Data: UNCOMTRADE database \cite{UNCOM}, year 2001 ($N=162$ countries). The values are rescaled by the total weight.}
\label{fig:weighted}

\end{figure}

Even when weighted properties of the network are studied, the importance of the underlying topology is still manifest, e.g. when local averages of weighted quantities are performed. 
For instance, let us consider the weighted analogue of the degree, i.e. the \emph{strength} (denoted by $s$) defined as the total weight of the links of a country (the total value of imports and exports for that country). 
As in the binary case, the average strength of the partners of a country (the \emph{average nearest neighbour strength}, denoted by $s^{nn}$) is found to decrease as the strength of that country increases \cite{mypre2,myjeic} (see fig. \ref{fig:weighted}).
Being a local average over the partners of each country, $s^{nn}$ is strongly influenced by the degree, which is a binary property.
Similarly, a weighted generalization of the clustering coefficient (denoted by $c^w$) is also influenced by the binary structure, since it is still defined on the local neighbourhood of countries.
Unlike its binary counterpart, $c^w$ is found to increase as the strength increases \cite{mypre2,myjeic} (see fig. \ref{fig:weighted}).

\section{Statistical physics and maximum-entropy models}\label{sec:maxent}
Taken together, the above findings highlight that the topology of the ITN is nontrivial and very different from the complete network predicted by the Gravity Model. 
In the previous section we discussed some coincidences and deeper similarities behind the use of the gravity law in physics and economics.
As another interesting similarity, recent results \cite{mypre1,mypre2,myjeic,rossana} suggest that the limitations of the Gravity Model can only be overcome after a change of paradigm which is not dissimilar from the one that accompanied two revolutions in physics, namely the advent of statistical mechanics and that of quantum physics.
The new paradigm assumes that, in order to predict the presence of a link (and not only its weight), \emph{probabilistic models} of networks need to be considered.
The great conceptual jump consists in assuming that the observed network is not unique, but one of many possible realizations, each of which has a probability $P$ to occur.
This probability must be determined by establishing which of the properties of the network are somehow `unavoidable', and assuming that all possible networks displaying those properties (including the real network) are equally probable.
This change of approach is equivalent to the one leading to the introduction of statistical physics: if the detailed \emph{microscopic} configuration of a large system is unknown (as is generally the case), and only a few \emph{macroscopic} quantities are known (e.g. the total energy), then some properties of the system can be inferred by averaging over all possible configurations compatible with the known macroscopic quantities.
The probability of each configuration therefore depends on the choice of the macroscopic quantities to be reproduced.

As Jaynes pointed out in his work devoted to the connections between statistical mechanics and information theory \cite{jaynes}, the fundamental problem of statistical physics can be regarded as a particular case of a more general class of problems of inference from partial information.
In the general case, one looks for the probability distribution that maximizes the uncertainty about the system, given the partial knowledge of the latter.
Mathematically, if $C$ denotes a possible (microscopic) configuration of the system, the solution of the problem is obtained maximizing Shannon's entropy
\begin{equation}
S\equiv-\sum_C P(C)\ln P(C)
\label{eq:S}
\end{equation}
subject to a set of \emph{constraints}, representing what is known about the system \cite{jaynes}. 
The result of this constrained maximization problem is the probability
\begin{equation}
P(C)=\frac{e^{-H(C)}}{Z}
\label{eq:pc}
\end{equation}
where $H(C)$ is a linear combination of the constraints (where each constraints is coupled to its Lagrange multiplier) and
\begin{equation}
Z\equiv \sum_C e^{-H(C)}
\label{eq:Z}
\end{equation}
is the normalization factor.
The ensemble of configurations generated by eq.(\ref{eq:pc}) is the \emph{maximum-entropy ensemble} specified by the chosen constraints.

Jaynes noticed that, if the system under consideration is a physical one, and the only constraint is the total (macroscopic) energy $E(C)$, then $H(C)=\beta E(C)$ where $\beta$ is the Lagrange multiplier ensuring that (the ensemble average of) $E(C)$ can be set equal to its observed value.
This leads to the identification of eq.(\ref{eq:pc}) with the Gibbs-Boltzmann distribution, if $\beta$ is identified with the inverse temperature through $\beta=(kT)^{-1}$ ($k$ being Boltzmann's constant).
Automatically, this also shows that eq.(\ref{eq:Z}) can be identified with the \emph{partition function}, and eq.(\ref{eq:S}) with the Gibbs-Boltzmann entropy.

Coming to the case of networks, it has been shown \cite{newman_expo,aga1,mymethod,aga2} that the class of network models known in the social sciences under the name of \emph{Exponential Random Graphs} or $p^*$ models \cite{newman_expo} is also a particular case of the above maximum-entropy problem.
In these models, the constraints are (not necessarily macroscopic) topological properties that one wants to control for. 
Consider for instance the case of \emph{binary graphs} with a given number $N$ of vertices.
In undirected binary graphs, each pair of vertices is either connected or not, with no possible variation in the direction and intensity of the connection.
Each configuration $C$ is uniquely specified by the \emph{adjacency matrix} $A$ of the graph, defined as a symmetric $N\times N$ matrix with entries ${a_{ij}=1}$ if a link exists between the vertices $i$ and $j$, and ${a_{ij}=0}$ otherwise.
Therefore, we can label each configuration with $A$ rather than $C$, and the corresponding probability with $P(A)$.
The simplest example is when the only constraint is the total number $L$ of links. It has been shown \cite{newman_expo,mymethod} that this particular case reduces to the popular Erd\H{o}s-R\'enyi random graph model, where all pairs of vertices are connected independently of each other and with the same probability $p$ (which can be viewed as the only Lagrange multiplier, uniquely specified by the observed value of $L$).
In this model, eq.(\ref{eq:pc}) translates into the following probability $P(A)$, that simply factorizes over all pairs of vertices $i,j$: 
\begin{equation}
P(A)=\prod_{i<j}p^{a_{ij}}(1-p)^{1-a_{ij}}=p^{L(A)}(1-p)^{N(N-1)/2-L(A)}
\label{formula2.5}
\end{equation}
This expression shows that the generation of an entire graph $A$ is the combination of $N(N-1)/2$ independent \emph{Bernoulli trials}\footnote{\label{foot:bernoulli}A \emph{Bernoulli trial} (or \emph{Bernoulli process}) is the simplest random event, i.e. one characterized by only two possible outcomes. One of the two outcomes is referred to as the `success' (in this case, the creation of a link) and is assigned a probability $p$. The other outcome is referred to as the `failure', and is assigned the complementary probability $1-p$. Equation (\ref{formula2.5}) is indeed the product of the probability $p^{L(A)}$ of $L(A)$ successful events of link creation times the probability $(1-p)^{N(N-1)/2-L(A)}$ of the complementary number of failures, where $L(A)$ is the number of links in the particular graph $A$. Note that $N(N-1)/2$ is the total number of pairs of $N$ vertices: we are uninterested in self-loops, so the diagonal matrix entries are $a_{ii}=0$, which leaves us with only $N(N-1)/2$ degrees of freedom in a symmetric $N\times N$ adjacency matrix. For the same reason, the sum in eq.(\ref{formula2.5}) runs over pairs with $i<j$, i.e. only over the upper triangle of the matrix $A$.}, each corresponding to the creation of a single link and characterized by the same success probability $p$. 

The above simple example shows that individual links naturally inherit, from the maximum-entropy structure of the overall model, the character of random variables, to be described by probability distributions. 
If weighted networks are considered, maximum-entropy models can still be defined \cite{newman_expo,mymethod} and lead again to a probabilistic description of the weight of all links, \emph{including the possibility of zero weights which correspond to missing links} (we will discuss explicit cases later).
Thus, in both binary and weighted descriptions, the probabilistic character of link creation characterizing the maximum-entropy approach eliminates the need to specify the topology of the network as `given', overcoming the limitation encountered when using the Gravity Model.
This makes maximum-entropy graph ensembles a potentially successful approach to the analysis of the ITN and economic networks in general.
However, two aspects remain to be discussed:
\begin{itemize}
\item one needs of course to check whether a suitable choice of constraints can indeed reproduce the empirical properties of the ITN: this requires the identification of topological constraints that are both reasonable (i.e. they can be \emph{a priori} justified as a meaningful choice) and effective (i.e. they are \emph{a posteriori} successful in replicating the ITN);\\
\item even if the specification of appropriate \emph{topological} constraints turns out to satisfactorily reproduce the observed network, one needs to understand whether this result can be reconciled with, or at least related to, the main idea of the Gravity Model: the assumption that trade strongly depends on \emph{non-topological} quantities such as GDP and distances.
\end{itemize}
In sections \ref{sec:2}-\ref{sec:4} we will address the first point in detail, while in sec. \ref{sec:gdp} we will deal with the second one.

So far, our general discussion has highlighted that the transition from the Gravity Model to maximum-entropy models is analogous, both conceptually and mathematically, to the paradigm shift that led to the introduction of Statistical Physics at the beginning of the twentieth century. 
The common aspect in both cases is the \emph{probabilistic} description of the system.
As we now show, another common aspect leads to a further formal similarity: the \emph{discreteness} of both economic quantities and microscopic particles implies that, when a specific choice of constraints is made, eq.(\ref{eq:pc}) leads to the same mathematical expressions that are encountered in Quantum Physics. 
These expressions are the so-called \emph{Fermi-Dirac} and \emph{Bose-Einstein statistics}.\footnote{In quantum physics, fundamental particles are believed to be of two types: \emph{fermions} or \emph{bosons}, depending on the value of their \emph{spin} (an intrinsic `angular moment' of the particle). Fermions have half-integer spin and cannot occupy a quantum state (a configuration with specified microscopic degrees of freedom, or \emph{quantum numbers}) that is already occupied. In other words, at most one fermion at a time can occupy one quantum state. The resulting probability that a quantum state is occupied is known as the \emph{Fermi-Dirac} statistics. Bosons have integer spin and can occupy states with no restriction: any non-negative integer number of bosons can occupy the same quantum state. The resulting expected number of particles occupying a given quantum state is described by the so-called \emph{Bose-Einstein} statistics.\label{foot:quantum}} 

\section{Fermi-Dirac statistics}
\label{sec:2}
Let us first consider binary networks, i.e. let us focus only on the presence/absence of links. In order to simplify the discussion, let us also consider undirected networks (the results that follow can be straighforwardly extended to directed configurations). 
We have already mentioned the example of the random graph model, obtained when the only constraint is the total number of links.
That model is very simple, but severely limited by its complete homogeneity: all vertices have approximately the same topological properties, narrowly distributed around a common average value. 
This is in stark contrast with the strong heterogeneity of most real-world economic networks, including the ITN as we already discussed in section \ref{sec:network}.
If we want to build a maximum-entropy model of the ITN whose topology is a real improvement over the Gravity Model, we need to reproduce the observed heterogeneity of the network. 
To this end, it is necessary to enforce different constraints that lead to more complicated models.
One of the widespread choices in network theory is to consider an ensemble of networks where each vertex $i$ has the same degree $k_i$ as in the real network. 
This choice is justified by the fact that, being an entirely local topological property, the degree is expected to be directly affected by some intrinsic (non-topological) property of vertices. 
For instance, we already anticipated that in the ITN the degree of a country increases with the GDP of the latter \cite{mywtw}.
It would of course not make sense to compare the real ITN with a randomized counterpart where the degree of a country no longer corresponds to a realistic value of its GDP (for instance, where the USA have only one or two connections).
This leads us to interpret the observed degrees of countries as `unavoidable' topological constraints, in the sense that the violation of the observed values would lead to an `impossible', or at least very unrealistic, world trade network.

The resulting model is known as the \emph{Configuration Model}, and is defined as a maximum-entropy ensemble of graphs with given \emph{degree sequence} \cite{newman_expo,mymethod}. The degree sequence, which is the constraint defining the model, is nothing but the ordered vector $\vec{k}$ of degrees of all vertices (where the $i$th component $k_i$ is the degree of vertex $i$).
The ordering preserves the `identity' of vertices: in the resulting network ensemble, the expected degree $\langle k_i\rangle$ of each vertex $i$ is the same as the empirical value $k_i$ for that vertex.
In the Configuration Model, eq.(\ref{eq:pc}) translates into the graph probability
\begin{equation}
P(A)=\prod_{i<j}q_{ij}(a_{ij})=\prod_{i<j}p_{ij}^{a_{ij}}(1-p_{ij})^{1-a_{ij}}
\label{formula3}
\end{equation}
where $q_{ij}(a)=p_{ij}^{a}(1-p_{ij})^{1-a}$ is the probability that that particular entry of the adjacency matrix $A$ takes the value $a_{ij}=a$. 
The above expression shows that the creation of a link has still the form of a Bernoulli process (see footnote \ref{foot:bernoulli}), but now, unlike the random graph model described by eq.(\ref{formula2.5}), different pairs of vertices are characterized by different connection probabilities $p_{ij}$. 
These probabilities read \cite{mymethod}
\begin{equation}
\langle a_{ij}\rangle=p_{ij}
=\frac{x_{i}x_{j}}{1+x_{i}x_{j}}
\label{formula4}
\end{equation}
where $x_i$ is the Lagrange multiplier obtained by ensuring that the expected degree of the corresponding vertex $i$ equals its observed value: $\langle k_i\rangle=k_i$ $\forall i$ \cite{mymethod}.
Note that, as always happens in maximum-entropy ensembles  described by eq.(\ref{eq:pc}), the probabilistic nature of configurations implies that the constraints are valid only on average (the angular brackets indicate an average over the ensemble of realizable networks).
Also note that $p_{ij}$ is a monotonically increasing function of $x_i$ and $x_j$.
This implies that $\langle k_i\rangle$ is a monotonically increasing function of $x_i$.
An important consequence is that two countries $i$ and $j$ with the same degree $k_i=k_j$ must have the same value $x_i=x_j$.

Equation (\ref{formula4}) provides an interesting connection with quantum physics, and in particular the statistical mechanics of the microscopic particles known as \emph{fermions} (see footnote \ref{foot:quantum}).
The `selection rules' of fermions dictate that only one particle at a time can occupy a single-particle state, exactly as each pair of vertices in binary networks can be either connected or disconnected. In this analogy, every pair $i,j$ of vertices is a `quantum state' identified by the `quantum numbers' $i$ and $j$.
So each link of a binary network is like a fermion that can be in one of the available states, provided that no two objects are in the same state. 
Equation (\ref{formula4}) indicates the expected number of particles/links in the state specified by $i$ and $j$.
With no surprise, it has the same form of the so-called \emph{Fermi-Dirac statistics} describing the expected number of fermions in a given quantum state \cite{newman_expo,mymethod,aga2}. 
As we already discussed, the probabilistic nature of links allows also for the presence of empty states, whose occurrence is now regulated by the 
probability coefficients $(1-p_{ij})$.

\begin{figure}[t!]
%\sidecaption
\includegraphics[scale=.38]{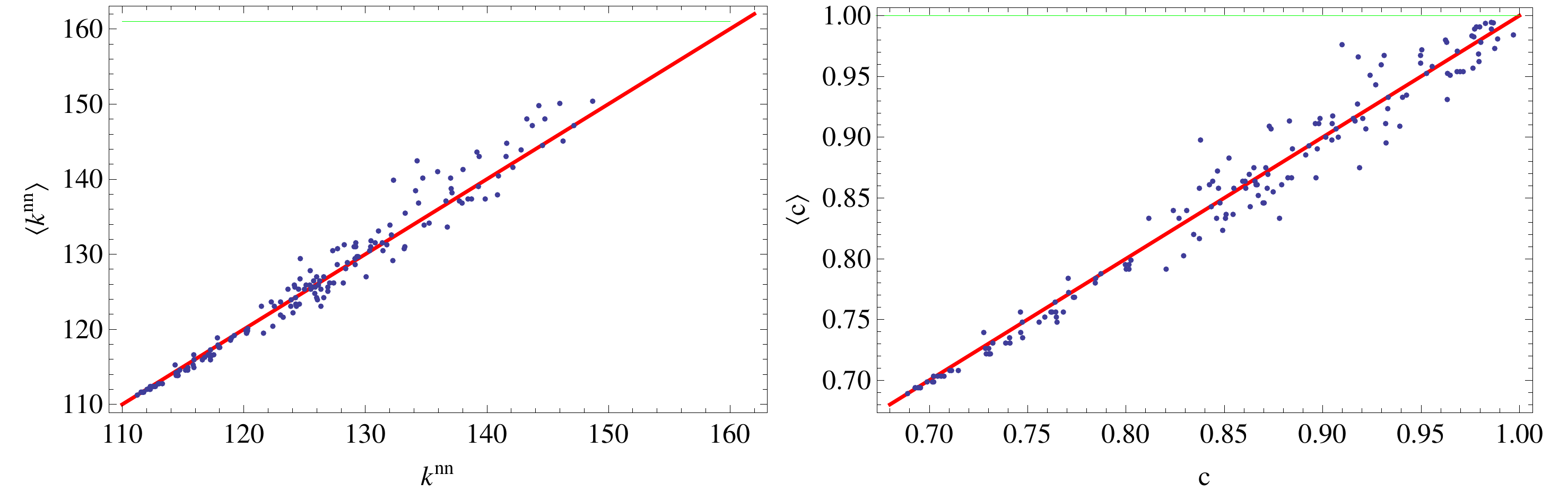}
\caption{The topology of the World Trade Web is well reproduced by specifying the number of trade partners of each country (binary Configuration Model). Left: observed VS expected average nearest neighbour degree $k^{nn}$, for all vertices. Right: observed VS expected clustering coefficient $c$. 
The red curves are identity lines (perfect agreement).
The green curves represent the prediction for the same quantities under the Gravity Model, which instead predicts a completely connected network with all vertices characterized by the same value. Data: UNCOMTRADE database \cite{UNCOM}, year 2001 ($N=162$ countries).}
\label{fig:1}
\end{figure}

We now come to the application of the model to the topology of the ITN.
Unlike the Gravity Model, the Configuration Model allows the whole degree sequence of the observed network to be preserved (on average), while randomizing other (unconstrained) network properties \cite{mymethod}. 
In order to check whether the model successfully reproduces the ITN, one needs to compare the higher-order (unconstrained) observed topological properties with their expected values calculated over the maximum-entropy ensemble.
This automatically indicates whether the degree sequence is informative in explaining the rest of the topology.
This can be done analytically, by means of the 
probabilities appearing in eq. (\ref{formula4}) \cite{mymethod}.
The effectiveness of the degree sequence in reproducing other topological properties of the ITN is shown in fig. \ref{fig:1}, where we compare the observed values of the average nearest neighbour degree $k^{nn}_i$ and clustering coefficient $c_i$ (defined in sec. \ref{sec:network}) with the corresponding expected values $\langle k^{nn}_i\rangle$ and $\langle c_i\rangle$, for all vertices. 
In this type of plot, the agreement between model and observations can be simply assessed as follows: the less scattered the cloud of points around the identity function, the better the agreement between model and reality. 
In principle, a broadly scattered cloud around the identity function would indicate the little effectiveness of the chosen constraints in reproducing the unconstrained properties, signalling the presence of genuine higher-order patterns of self-organization, not simply explainable in terms of the degree sequence alone. 
However, the results in fig. \ref{fig:1} indicate that the World Trade Web is well reproduced by the Configuration Model.
This result is very robust, as documented by recent analyses that have confirmed it for different temporal snapshots, different levels of aggregation (up to individual commodities), and different datasets \cite{mypre1,myjeic}. 
With the appropriate generalizations, this conclusion also holds when the ITN is analysed as a \emph{directed} network, still well described by the Fermi-Dirac statistics  \cite{mypre1,mytriadic}.

For completeness, fig. \ref{fig:1} also shows the Gravity Model's prediction of a complete network, which is dramatically different from the observed one. 
Thus the maximum-entropy approach, and in particular the Configuration Model, represents a significant advantage with respect to the Gravity Model.
An unexpected implication is that the degrees of world countries are maximally informative about the ITN as a whole: if the empirical degree sequence of the ITN is not reproduced, the observed topology of the network as a whole will not be reproduced either \cite{mypre1}. 
Unfortunately, current micro-founded models in the economics literature do not attempt at replicating or explaining the particular value of the number of trade partners of a country. 
Rather, they aim at reproducing the Gravity Model, inspired by the success of the latter at the level of non-zero flows \cite{anderson}.
In so doing, these models are destined to fail in explaining the heterogeneous topology of the ITN.
To overcome this limitation, future models should aim at replicating the degree sequence explicitly.
As we anticipated at the end of sec. \ref{sec:maxent}, one should also investigate the relationship between the `empirically informative constraints' in the maximum-entropy approach (the degree sequence in this case) and the `macroeconomic explanatory factors' in the Gravity Model approach  (such as GPD and distances).
We will discuss this important point in sec. \ref{sec:gdp}.

We conclude this section by stressing again that the `fermionic' character of the ITN, when treated as a binary network, is the mere result of the restriction that no two binary links can be placed between any two vertices, leading to a mathematical result which is formally equivalent to the one of quantum statistics. Clearly, there is nothing really `quantum' in trade connections being described by the Fermi-Dirac statistics, exactly as there is nothing really `gravitational' in non-zero trade flows being described by the Gravity Model.
Still, the deep epistemological analogies leading to similar laws in physics and economics (the ones we discussed in sec. \ref{sec:gravity}) remain, and are now translated into a more sophisticated formalism that allows for the probabilistic and discrete nature of the system:
\begin{itemize}
\item In both physical and economic applications, the Fermi-Dirac statistics has the following \emph{symmetry}: the probability $p_{ij}$ only depends on the combination $x_i x_j$. In quantum physics, $x_i x_j$ in turn depends only on the energy of the quantum state, while in the Configuration Model it depends only on the end-point degrees $k_i$ and $k_j$. 
This means that, all else being equal (e.g. given the same energy, or the same values of the end-point degrees), the occupation probability of two different states $(i,j)$ and $(m,n)$ is the same.\\
\item In both applications, the Fermi-Dirac statistics implements the \emph{qualitative} idea that, the larger the value of $x_i x_j$, the higher the probability that the state $(i,j)$ is occupied.\\
\item Finally, the validity of the Fermi-Dirac statistics is in both cases established \emph{a posteriori}, by the fact that it reproduces empirical observations.
This phenomenological agreement confirms that the postulated quantum numbers/topological constraints, which uniquely specify the values $\{x_i\}$, are indeed the (only) relevant ones for the problem under investigation.
\end{itemize}
As we will discuss in sec. \ref{sec:gdp}, in the ITN the value of $x_i x_j$ can also be related to the GDP of the two countries $i$ and $j$, and to the geographic distance separating them. This restores a tight correspondence between the three points listed above and the three ones discussed in section \ref{sec:gravity}.

\section{Bose-Einstein statistics}
\label{sec:3}
We started this chapter stressing the importance of the ITN as a complex weighted network, while in the previous section we restricted ourselves to the description of its purely binary topology. 
From this point onwards, we go back to the full weighted level and discuss whether it is possible to reproduce the topology and weights of the ITN simultaneosly.
Naively, the results shown so far suggest that a first attempt in this direction could be the introduction of a two-step process where the topology is first established using the Configuration Model, and the realized link weights are then estimated using the Gravity Model.
However, besides being disappointingly inelegant, this approach would result in a hybrid combination of maximum-entropy and economically inspired expectations, leaving the final results without a clear interpretation.
A more satisfactory way to proceed is expanding the maximum-entropy formalism into one valid for weighted networks, automatically closer to the macroeconomic reasoning.

To many economists, the finding that the number of trade partners of a country is a particularly informative quantity might appear weak or misleading, given the expectation that the monetary value of imports and exports is in principle much more informative:
common sense suggests that knowing how much (in dollars) a country trades with the rest of the world should be more informative than just knowing how many partners trade with that country.
This leads to the expectation that the strength $s$ should be more informative than the degree $k$.
One might therefore suspect that an even better model of the ITN, still preserving the observed heterogeneity of countries, is one where the strengths, rather than the degrees, are enforced as constraints.
In this section, we take this approach and show that it actually leads to a counter-intuitive result: unlike what we found for the degree sequence, knowing the strength of each world country turns out to be very poorly informative about the structure of the ITN as a whole.

The theoretical framework introduced in the previous section allowed us to treat links as probabilistic entities in order to overcome the Gravity Model's prediction of a completely connected network. 
Since we were only interested in the prediction of the presence or absence of links, the selection rules were formally analogue to the fermionic ones: $a_{ij}=0,\:1$. However, the formalism can be generalized in order to analyze \emph{weighted networks} where links can have non-negative integer weights. 
If we keep considering undirected graphs for simplicity, each network is now specified by a $N\times N$ symmetric \emph{weight matrix} $W$ whose entry $w_{ij}$ equals the weight of the link between the vertices $i$ and $j$.
Therefore, we can now label each configuration with $W$, and the corresponding probability with $P(W)$.
If we define the \emph{strength sequence} $\vec{s}$ as the ordered vector of strength values (with components $s_i$, $i=1,\dots,N$), the \emph{Weighted Configuration Model} can be introduced as the ensemble of \emph{weighted} networks with given strength sequence.
If one allows each $w_{ij}$ to take non-negative integer numbers, eq.(\ref{eq:pc}) now becomes \cite{mymethod}
\begin{equation}
P(W)=\prod_{i<j}q_{ij}(w_{ij})=\prod_{i<j}p_{ij}^{w_{ij}}(1-p_{ij}).
\label{formula5}
\end{equation}
Where $q_{ij}(w)=p_{ij}^{w}(1-p_{ij})$, which has now the form of a \emph{geometric distribution}\footnote{\label{foot:geometric}In one of its possible formulations, the \emph{geometric distribution} describes the probability that, in a sequence of repeated Bernoulli trials (see footnote \ref{foot:bernoulli}) with success probability $p$, the first $w$ trials are all successful and the following one is unsuccessful. This happens with probability $p^{w}(1-p)$.}, is the probability that the vertices $i$ and $j$ are connected by a link of weight $w$. The outcome $w=0$, corresponding to a missing link, occurs with probability $1-p_{ij}$.
Therefore $p_{ij}$ still denotes the probability that $i$ and $j$ are connected, irrespective of the weight of this connection.
This probability now reads 
\begin{equation}
p_{ij}=y_i y_j
\label{eq:pbose}
\end{equation}
where $y_i$ is the Lagrange multiplier required in order to ensure that the expected strength $\langle s_i\rangle$ of each vertex $i$ equals the empirical value $s_i$ \cite{mymethod}.
This time, two countries $i$ and $j$ with the same strength $s_i=s_j$ (independently of their degrees) must have the same value $y_i=y_j$ \cite{mypre2}.

As in the previous case, a connection with another well-known quantum statistics emerge. The `selection rules' have now allowed us to treat link weights as formally analogue to \emph{bosons} (see footnote \ref{foot:quantum}),
admitting multiple and unlimited occupations of the same `quantum state' ($w_{ij}=0,\:1,\:2\dots{+\infty}$).
Indeed, the expected occupation number of a quantum state, which is the expected weight of the link between vertices $i$ and $j$, is now formally identical to the so-called \emph{Bose-Einstein statistics} \cite{newman_expo,mymethod}:
\begin{equation}
\langle w_{ij}\rangle=\frac{p_{ij}}{1-p_{ij}}
=\frac{y_{i}y_j}{1-y_{i}y_j}
\label{formula6}
\end{equation}
As before, there is nothing fundamental in the mathematical analogy with quantum statistics, the only common element being the postulated \emph{discreteness} of the numbers $w_{ij}$.\footnote{In quantum physics, the discreteness is implied by the fact that particles can only exist in integer number. In economic networks, the discreteness is implied by the fact that money can only exist in integer multiples of a fundamental, indivisible unit of currency (such as one Eurocent).}
The deeper similarity involves again the concept of  symmetry, which in this case refers to the assumption that, all else being equal, in the Bose-Einstein statistics the expected value $\langle w_{ij}\rangle$ only depends on $y_i y_j$. 
Similarly, the common qualitative aspect shared by the physical and economic applications is that $\langle w_{ij}\rangle$ is expected to increase with $y_i y_j$.
However, as we now show, this time the phenomenological analogy between the Weighted Configuration Model of trade and quantum statistics (i.e. the agreement of both with real data) breaks down:
while the Bose-Einstein distribution describes the microscopic world of bosons remarkably well, eqs.(\ref{formula5}-\ref{formula6}) fail miserably in reproducing the observed ITN.

\begin{figure}[t!]
%\sidecaption
\includegraphics[scale=.38]{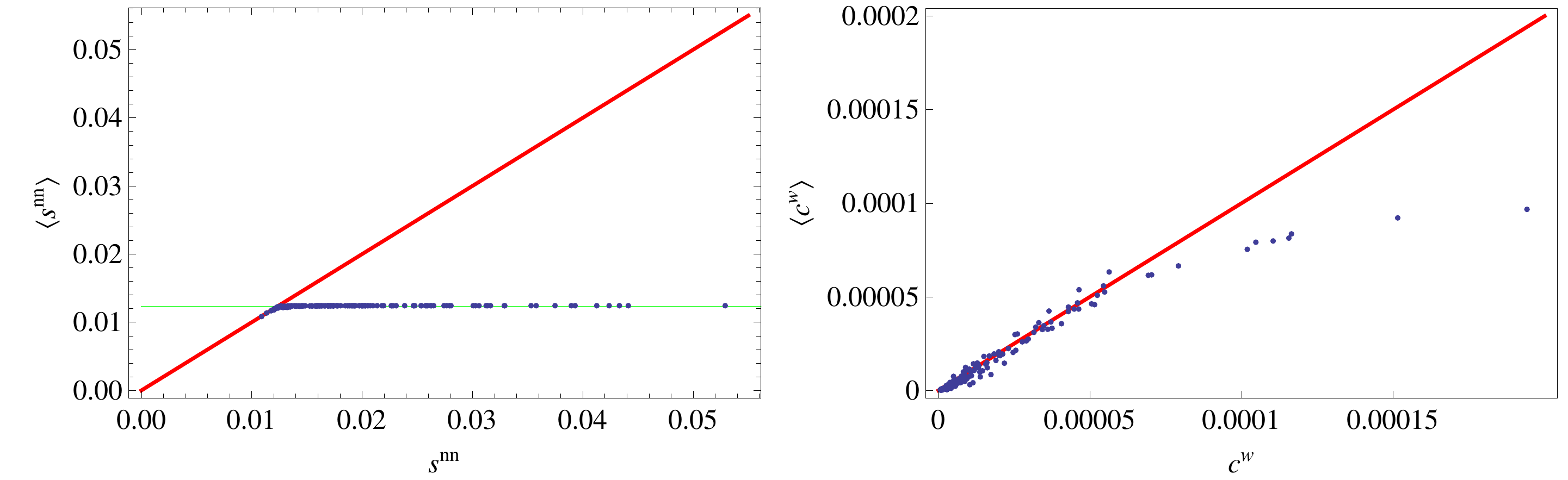}
\caption{The topology and weights of the World Trade Web are NOT well reproduced by specifying the total import and export value of each country (weighted Configuration Model). Left: observed VS expected average nearest neighbour strength $s^{nn}$, for all vertices. Right: observed VS expected weighted clustering coefficient $c^w$. 
The red curves are identity lines (perfect agreement).
The green curve represents the prediction for the same quantities under the Gravity Model, which instead predicts a completely connected network with all vertices characterized by the same value. Data: UNCOMTRADE database \cite{UNCOM}, year 2001 ($N=162$ countries). The values are rescaled by the total weight.}
\label{fig:2}
\end{figure}

To see this, one can again compare the observed and the expected values of higher-order (unconstrained) topological properties. 
In fig. \ref{fig:2} we show the average nearest neighbour strength $s^{nn}$ and weighted clustering coefficient $c^w$ (see sec. \ref{sec:network}).
The results now indicate how bad the accordance between the Weighted Configuration Model and the real network is. 
Interestingly enough, the model's prediction for the average nearest neighbors strength is almost identical to the Gravity Model's prediction for the same quantity. 
This unambiguously indicates that the two models suffer from the same limitation: their incapabaility to reproduce the topology of the observed network and, in particular, the fact that the Weighted Configuration Model generates an extremely dense network \cite{mypre2}, not too different from the completely connected topology predicted by the Gravity Model (for a fully connected network whose weights are rescaled by the total weight, it is easy to estimate $s^{nn}\simeq \sum_{i}s_i/N=2/N$). 
Similarly, even if the smaller values of the weighted clustering coefficient seem to partially agree with the model's prediction, the behaviour for large values indicates that major refinements are needed in order to improve the model performance \cite{mypre2}.
The disagreement between the Weighted Configuration Model and the real ITN has been confirmed on different data, different temporal snapshots, and different commodities \cite{mypre2,myjeic}.

The above result contradicts the intuitive expectation that, by taking a weighted quantity (the strength sequence $\vec{s}$) as input, the Weighted Configuration Model should be more informative than its binary counterpart.
Indeed, while the \emph{complete} knowledge of all the weights of the network is of course more informative than the knowledge of the binary topology alone, it turns out that the \emph{partial} knowledge of the weighted network (the strength sequence in this case) is \emph{less} informative than the knowledge of the corresponding binary quantity (the degree sequence).
A somewhat puzzling consequence for macroeconomic modeling is that, even if a micro-founded model of international trade successfully reproduces the observed total imports and exports of all world countries, this is definitely not enough in order to reproduce the structure of the ITN as a whole.
If combined with the previous result about the extreme informativeness of the degree sequence, this finding strengthens the unconventional conclusion that \emph{satisfactory models of international trade should aim at primarily reproducing the binary properties of countries (number of trade partners) rather than their weighted ones (total import and export)} \cite{mypre1,mypre2,myjeic}.

\section{Generalized quantum distributions}
One therefore needs to look for a better model of the International Trade Network, able to reproduce the binary topology and the weighted structure of the network simultaneously.
Since microscopic particles are either fermions or bosons, the Fermi-Dirac and Bose-Einstein distributions are the only two types of quantum statistics traditionally used in physics.\footnote{Excluding the case of \emph{anyons} \cite{anyons}, particles that can only exist in two dimensions and that can be described by a generalized `fractional statistics' \cite{fracstat}, which is however unrelated to the extensions we discuss in this section and in the next one.}
However, one can formally introduce generalized distributions that reduce to the Fermi-Dirac and the Bose-Einstein statistics as particular cases.
While the physically realizable quantum systems might correspond only to the fermionic and bosonic extremes, it might well be that other systems, such as economic networks, can instead realize other non-trivial limits of those generalized distributions.
Therefore, in this and in the next section we discuss two possible generalized `quantum' statistics and their relationship with the structure of economic networks, and the ITN in particular.

The case we consider in this section is just a pedagogical example, while the one we discuss in the next section leads to a very important model that reproduces the observed ITN in great detail.
As we showed, the only mathematical ingredient needed to generate the Fermi-Dirac statistics in maximum-entropy network ensembles is the $0/1$ character of binary links ($a_{ij}=0,1$), while the only ingredient needed to generated the Bose-Einstein statistics is the non-negative integer character of weighted links ($w_{ij}=0,\dots,+\infty$).
For non-physical systems, there is no reason why the only two allowed values for the maximum weight should be one and infinite. 
In general, we can consider a general family of distributions, obtained when the occupation number can range from $0$ to a \emph{finite} maximum value $w_{max}$. 
All the distributions within this family share the same discrete character, due to the integer occupation numbers of `quantum states'. 
However, they only reduce to the Fermi-Dirac and Bose-Einstein distributions in the extreme cases ${w_{max}=1}$ and ${w_{max}\to+\infty}$ respectively.

\begin{figure}[t!]
\sidecaption
\includegraphics[scale=0.48]{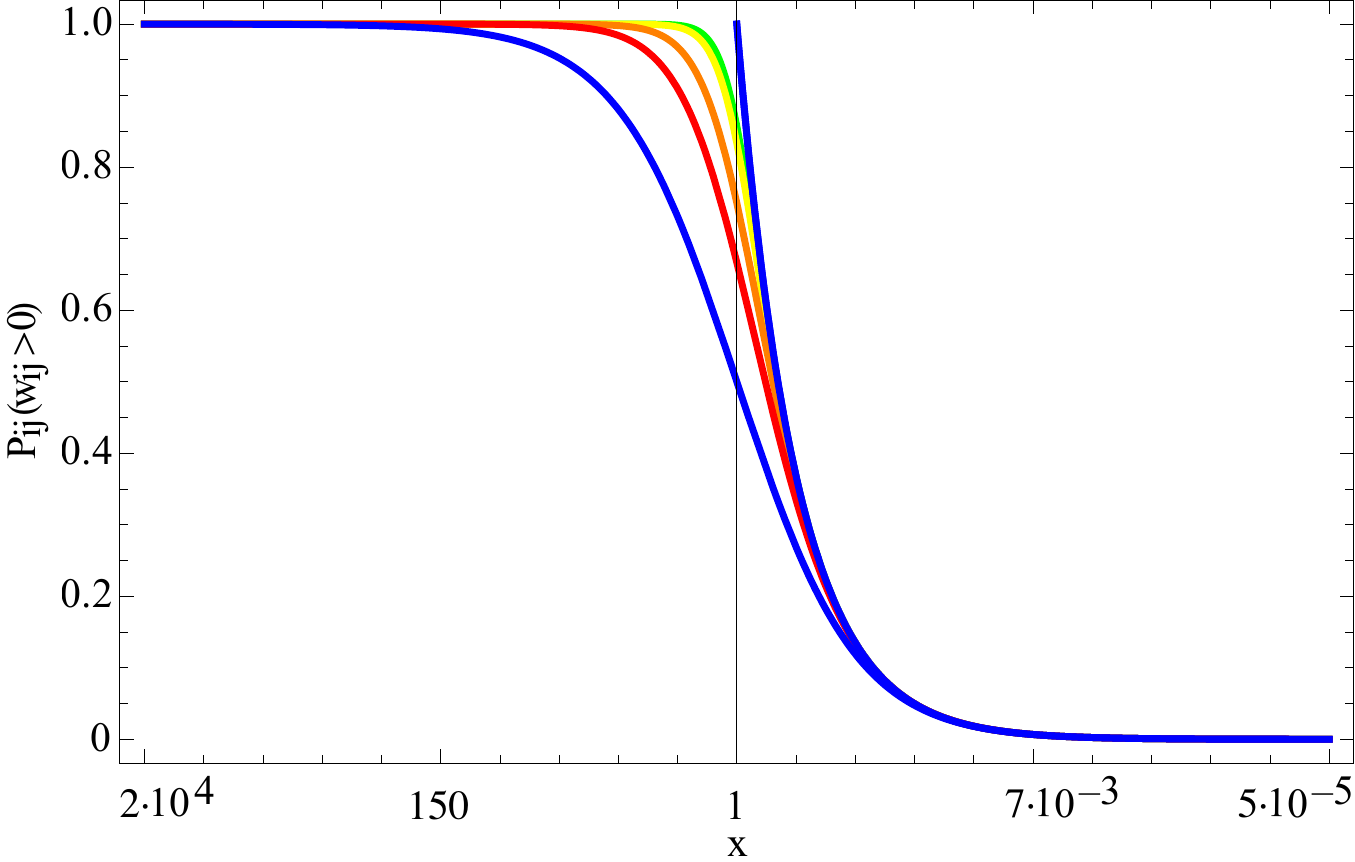}
\caption{Single-link cumulative distribution functions of the generalized statistics, defined in eq.(\ref{eq:cumul}), 
ranging from the Fermi-Dirac (the blue one on the left, with ${w_{max}=1}$) to the Bose-Einstein (the blue one on the right, with ${w_{max}\to\infty}$). The $x$-axis is logarithmically scaled.}
\label{fig:3}
\end{figure}

In the general case, with $w_{ij}=0,\:1,\:2\dots w_{max}$, the maximum-entropy ensemble of networks with given strength sequence $\vec{s}$ is characterized by the probability distribution
\begin{equation}
P(W)=\prod_{i<j}q_{ij}(w_{ij})=\prod_{i<j}\left[\frac{y_{ij}(1-y_{ij})}{1-y_{ij}^{(w_{max}+1)}}\right]^{w_{ij}}\left[\frac{1-y_{ij}}{1-y_{ij}^{(w_{max}+1)}}\right]^{1-w_{ij}}
\label{formula:7}
\end{equation}
where for simplicity we have defined $y_{ij}\equiv y_i y_j$, if $y_i$ still denotes the Lagrange multiplier needed to enforce the constraint $\langle s_i\rangle=s_i$. 
Like eqs. (\ref{formula3}) and (\ref{formula5}), the above probability is still a product over single-link distributions, each characterized by the same, bounded range of values. 
In order to better visualize the functional form of such distributions, the corresponding single-link cumulative distribution functions can be plotted, as shown in fig. \ref{fig:3}. The latter can be computed quite easily as
\begin{equation}
P_{ij}(w_{ij}>0)\equiv1-q_{ij}(0)=\frac{y_{ij}(1-y_{ij}^{w_{max}})}{1-y_{ij}^{(w_{max}+1)}}.
\label{eq:cumul}
\end{equation}
As $w_{max}$ increases from $1$ to $+\infty$, the intersections of these distributions with the $y$-axis form an interesting numerical succession, whose generic term is \begin{equation}
P_{ij}(w_{ij}>0|y_{ij}=1)=\frac{w_{max}}{w_{max}+1}=\frac{1}{2},\:\:\frac{2}{3},\:\:\frac{3}{4}\dots
\end{equation}
and whose limit when ${w_{max}\to+\infty}$ is $1$.

The generalized distribution considered above is an example showing how it is possible to gradually interpolate between the Fermi-Dirac and Bose-Einstein through the introduction of an extra parameter ($w_{max}$).
In principle, intermediate values of the parameter can lead to different results than the ones we showed in sec. \ref{sec:3}, and potentially to an improvement over them. However, since the observed weights in the ITN are extremely large, the value of $w_{max}$ required in order to generate realistic weights will be so large that the results are practically indistinguishable from those we have already discussed using the Bose-Einstein distribution.
So this model does not represent a real improvement.

\section{Mixed Bose-Fermi statistics}
\label{sec:4}
A fundamentally different way to unify the Fermi-Dirac and Bose-Einstein distributions, other than introducing an extra parameter while keeping the strength sequence $\vec{s}$ as the constraint, is adopting a different choice of the constraint themselves. 
Specifically, motivated by the success of the binary Configuration Model discussed in sec. \ref{sec:3}, one can introduce a maximum-entropy ensemble of networks with given degree sequence $\vec{k}$ and strength sequence $\vec{s}$, i.e. one where both constraints are enforced simultaneously \cite{mybosefermi}.
To this end, the maximum allowed value of weight is again $w_{max}=+\infty$ as in the Bose-Einstein case.
However, in terms of theoretical models, this leads to a whole new family of probability distributions, whose functional form is
\cite{mybosefermi}
\begin{equation}
P(W)=\prod_{i<j}q_{ij}(w_{ij})=\prod_{i<j}\left[\frac{(x_{i}x_{j})^{a_{ij}}y_{i}y_{j}(y_{i}y_{j})^{w_{ij}-1}(1-y_{i}y_{j})}{1-y_{i}y_{j}+x_{i}x_{j}y_{i}y_{j}}\right]
\label{eq_mix}
\label{formula:9}
\end{equation}
where $a_{ij}$, the element of the adjacency matrix, is $0$ if $w_{ij}=0$ and $1$ if $w_{ij}>0$. 
In the above expression, the $\vec{x}$ vector controls for the degrees and the $\vec{y}$ vector controls for the strengths.
This double set of constraints implies that, while different pairs of vertices are still independent, the creation of a link of given weight between two vertices is neither a Bernoulli nor a geometric process (see footnotes \ref{foot:bernoulli} and \ref{foot:geometric}), but a combination of the two.

As in the previous example, the (now generalized) `quantum' or discrete character of the statistics becomes evident as soon as the expected occupation numbers are computed \cite{mybosefermi}:
\begin{equation}
\langle a_{ij}\rangle=\frac{x_ix_jy_iy_j}{1-y_iy_j+x_ix_jy_iy_j}\quad\langle w_{ij}\rangle=\frac{x_ix_jy_iy_j}{(1-y_iy_j+x_ix_jy_iy_j)(1-y_jy_j)}
\label{eq:bosefermi}
\end{equation}
In this case, the unification of the Fermi-Dirac and Bose-Einstein distributions is achieved by combining binary and weighted constraints `as a block', i.e. in a single big step.
As a result, one cannot gradually interpolate between the two ordinary statistics: for instance, in order to retrieve the Bose-Einstein distribution, one must drop the entire degree sequence $\vec{k}$ from the set of constraints (mathematically, this corresponds to set $\vec{x}$ equal to the unit vector $\vec{1}$).\footnote{Note that the dual operation, i.e. dropping the entire strength sequence $\vec{s}$ from the set of constraints (mathematically corresponding to $\vec{y}=\vec{1}$), leads to an undefined model and does not correspond to the Fermi-Dirac statistics. The reason is that the maximum weight is still $w_{max}=+\infty$ and not $w_{max}=1$: without constraints on the weighted properties, the expected weights become infinite.}
The fundamental difference with respect to the intermediate distributions defined by eq. (\ref{formula:7}) is that now the occupation probability of an empty `single-link state' differs from the occupation probability of an already occupied state. 
In fact, this family of distributions can be intuitively described by saying that the `first' appearance of a link of unit weight between two disconnected vertices is regulated by the Fermi-Dirac statistics, while the `subsequent' appearance of units of weights between two already connected vertices is regulated by the Bose-Einstein statistics. For this reason, the statistics defined by eq.(\ref{eq:bosefermi}) is called the \emph{mixed Bose-Fermi statistics} \cite{mybosefermi}.

\begin{figure}[t!]
%\sidecaption
\includegraphics[scale=.39]{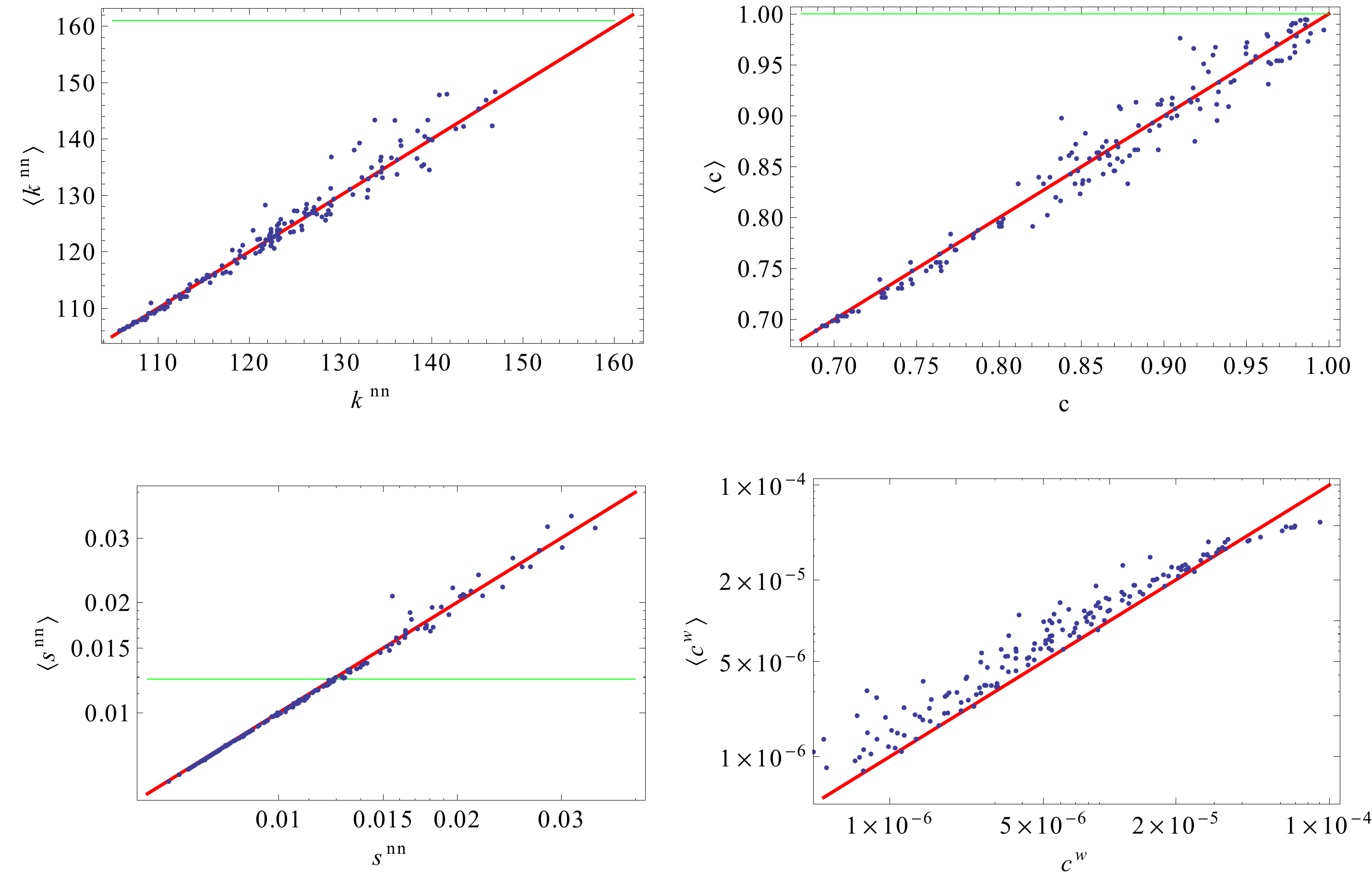}
\caption{The topology and weights of the World Trade Web are simultaneously well reproduced by specifying the number of partners of each country, as well as the total import and export value of each country (mixed Configuration Model). Top left: observed VS expected average nearest neighbour degree $k^{nn}$, for all vertices. Top right: observed VS expected clustering coefficient $c$. 
Bottom left: observed VS expected average nearest neighbors strength $s^{nn}$. Bottom right: observed VS expected clustering coefficient $c^w$. 
The red curves are identity lines (perfect agreement).
The green curves represent the prediction for the same quantities under the Gravity Model, which instead predicts a completely connected network with all vertices characterized by the same value. Data: UNCOMTRADE database \cite{UNCOM}, year 2001 ($N=162$ countries). The values are rescaled by the total weight.}
\label{fig:4}
\end{figure}

The Bose-Fermi statistics reproduces with great accuracy all the four higher-order structural quantities (both binary and weighted) of the ITN considered so far. 
This is shown in fig. \ref{fig:4}, where we compare the expected and observed values of $k_{nn}$, $c$, $s^{nn}$, and $c^w$ for all vertices. For the first time, we find a very close agreement for all these topological properties simultaneously. This result is very robust, as it holds for different snapshots and different commodities \cite{rossana}.
Two main conclusions can be drawn:
\begin{itemize}
\item on one hand, the addition of weighted constraints to the binary ones does not affect the effectiveness of the mixed model in reproducing the purely topological properties: the two upper panels of fig. \ref{fig:4} look approximately the same as the two panels of fig. \ref{fig:1};\\
\item on the other hand, the addition of binary constraints to the weighted ones dramatically improves the performance of the model in predicting purely weighted quantities. This is evident by comparing fig. \ref{fig:2} with the two bottom panels of fig. \ref{fig:4}. The latter show that now both the average nearest neighbour strength and the weighted clustering coefficient closely follow the identity function. In a sense, the addition of purely binary constraints compensates the incapability of the purely bosonic model in reproducing the network structure and brings the model back to high levels of performance at the topological level.
\end{itemize}
The family of mixed Bose-Fermi statistics not only represents a powerful model in order to explain \emph{both} the binary and the weighted quantities of interest of a given, observed network; it also points out the strong effects of the underlying topology on the weighted structural patterns. 

In economic terms, our discussion leads to the conclusion that the knowledge of monetary/weighted structural properties (such as total imports and exports) is informative only if in combination with non-monetary/binary properties (such as the number of trade partners).
The expectation that monetary quantities are \emph{per se} more informative than the corresponding binary ones turns out to be incorrect.  
For this reason, we believe that the Bose-Fermi statistics is a very useful tool in the understanding of economic networks in general. The fact that both strengths and degrees are enforced allows to study the interplay between the topological and monetary levels of organization, while still keeping the model parsimonious: only \emph{local} (country-specific) structural properties, the ones that we discussed as the somewhat irreducible and `unavoidable' level of heterogeneity, are enforced.

\section{The role of GDP and distance}
\label{sec:gdp}
The results presented so far show that the attempt to model the International Trade Network using analogies with physical laws, initiated by Jan Tinbergen with the introduction of the Gravity Model, turns out to be extremely successful, even if the appropriate formal expressions are very different from Tinbergen's original idea.
However, to complete our discussion, we need to address the final point anticipated in sec. \ref{sec:maxent}, i.e. how to reconcile maximum-entropy graph ensembles (that take structural properties as input) with the Gravity Model's expectation that international trade strongly depends on non-structural macroeconomic properties, such as GDP and geographic distances.

We start with a discussion about the role of GDP.
As we anticipated in sections \ref{sec:network} and \ref{sec:2}, the GDP of a country turns out to be highly correlated with the degree of that country in the ITN.
Interestingly, the functional dependence of the degree on the GDP can be adequately characterized by eq.(\ref{formula4}). 
In more detail, it was shown \cite{mywtw} that the connection probability 
\begin{equation}
p_{ij}=\frac{z\:GDP_i\:GDP_j}{1+z\:GDP_i\:GDP_j}
\label{eq:gdp}
\end{equation}
(where $z$ is a global free parameter) defines a model that reproduces the properties of the binary topology of the ITN very well, just like the Configuration Model defined by eq.(\ref{formula4}) does.
The value of $z$ is fitted to the data by requiring that the expected number of links $\langle L\rangle$ equals the observed number $L$ \cite{mywtw}.\footnote{This choice for the parameter $z$ corresponds to the maximization of the likelihood of the model defined by eq.(\ref{eq:gdp}) \cite{mylikelihood}, exactly like the values of $\{x_i\}$ that realize the conditions $\langle k_i\rangle=k_i$ $\forall i$ maximize the likelihood of the Configuration Model defined by eq.(\ref{formula4}) \cite{mymethod,mylikelihood}.}
This result shows that the parameter $x_i$ is approximately proportional to $GDP_i$, as can be confirmed explicitly \cite{mylikelihood}.
In terms of the network formation process, this means that the results discussed in sec. \ref{sec:2} can be almost entirely rephrased as follows. 
The GDP is found to determine \emph{directly} the number of trade partners of each country (because $GDP_i$ has the same role of the Lagrange multiplier $x_i$ determining $k_i$), and \emph{indirectly} the whole topology of the International Trade Network (because the functional form of the connection probability is that of the Configuration Model, where the higher-order topological properties are entirely determined by the degree sequence). The only topological quantity we need to know about the real network is the total number of links specifying the parameter $z$. 

In an only slightly more complicated way, it is also possible to incorporate distances into eq. (\ref{eq:gdp}) \cite{mydistances}. 
This leads to the probability
\begin{equation}
p_{ij}=\frac{z\:GDP_i\:GDP_j\:e^{-\gamma f(d_{ij})}}{1+z\:GDP_i\:GDP_j\:e^{-\gamma f(d_{ij})}}
\label{eq:distance}
\end{equation}
where $f(d_{ij})$ is some increasing function of the geographic distance between countries $i$ and $j$. The simplest choice for this function is $f(d_{ij})= d_{ij}$ \cite{myfilling}.
The model has now two parameters, which can be fixed simultaneously by imposing $\langle L\rangle=L$ and $\langle F\rangle =F$, where $F\equiv \sum_{ij}a_{ij}f(d_{ij})$ is a measure of the \emph{filling} of space by the network \cite{myfilling}.
A variant of this model has been recently used to analyse the directed version of the ITN \cite{mydistances}.  
The result one finds is that the addition of spatial information moderately improves the fit to the data.
However, alternative models that include information about the \emph{reciprocity} of trade \cite{mytriadic,mygrandcanonical}, rather than geographic distances, systematically outperform the spatial model \cite{mydistances}.

We note that, along the same lines as above, it is straightfoward to introduce GDPs and distances also in the weighted models defined by eq.(\ref{formula6}) and (\ref{eq:bosefermi}), by simply replacing ${y_i y_j}$ and ${x_i x_j}$ with ${z\:GDP_i\: GDP_j\: e^{-\gamma f(d_{ij})}}$.
Even if these weighted models have not yet been used in emprical analyses, the above discussion shows that maximum-entropy ensembles are not \emph{per se} incompatible with the Gravity Model's approach of explaining trade patterns in terms of macroeconomic quantities such as GDP and distances. 
On the contrary, we believe that maximum-entropy models are a very promising tool to understand economic networks. 
Identifying the most informative properties explaining the topology and weights of real economic networks is extremely important in order to identify the most relevant `targets' of theoretical models.
The finding that the observed trade patterns cannot be adequately understood unless one is able to reproduce the degree sequence, and that the latter is directly determined by the GDP, could only be established using a maximum-entropy model.
More in general, the important role played by binary properties even in weighted analyses is a highly nontrivial result.

\section{Conclusions}
In 1962, in what we would now call a pioneering attempt to model economic networks, the physicist and first Nobel Memorial Prize laureate Jan Tinbergen introduced the Gravity Model of trade mimicking Newton's gravitation law. 
This very intuitive and elegant proposal aimed at explaining trade exchanges in terms of a few macroeconomic quantities (GDP and geographic distance) by combining them in the same way as nature combines gravitational masses and spatial distances. 

The success of the Gravity Model is due to the fact that it reproduces well the observed (non-zero) trade flows between countries.
Minor refinements to the model, such as the inclusion of additional factors either favouring or suppressing trade, are relatively simple to make and further improve the fit to the data.
Therefore, for half a century the Gravity Model has been used more and more extensively in macroeconomic analyses, and it has become the standard model of international trade in the economics literature.
However, the most serious and in some sense irreducible limitation of the Gravity Model emerged only relatively recently, after the publication of several empirical analyses documenting the topology of the International Trade Network in the statistical physics literature.
While the Gravity Model predicts a completely connected network where every country trades with all other countries, the observed ITN is much more heterogeneous and hierarchical.

We have shown that this limitation can be overcome by adopting a probabilistic view, in exactly the same way as classical physics escaped its crisis at the end of XIX century by adopting the quantum-mechanical paradigm.
In network theory, this amounts to consider the adjacency matrix entries as probabilistic entities and the node pairs as single-link states whose occupation numbers are regulated by the same selection rules that apply to fermions and bosons in quantum physics. In this way, various probability distributions can be defined in order to explain the observed structural patterns. 
On one hand, purely fermionic selection rules excellently reproduce the binary topology of the ITN, but are intrinsically limited by the fact that they give no information about the weights of links in the network.
On the other hand, bosonic selection rules are suitable for weighted analyses but suffer from the same limitations of the Gravity Model, since they lead to the prediction of an almost completely connected network.
Interestingly enough, the most effective probabilistic models are those combining fermionic and bosonic selection rules. In this combination, the limitations encountered when the two quantum statistics are considered as separate are overcome simultaneously, and both the topology and weights of the observed ITN are nicely reproduced.

The main conclusion we can draw is the fundamental role played by topology in explaining the observed patterns of real world networks: in contrast with the `mainstream' economic thinking, purely weighted information (such as that encoded into the strength sequence) is \emph{not} enough to reproduce all the observed properties and, in particular, the purely binary ones (such as the degree sequence). A genuine, purely binary information is also needed from the very beginning, as confirmed by the successful family of mixed Bose-Fermi probabilistic distributions.
This shows that the naive expectation that weighted/monetary quantities are \emph{per se} more informative than the corresponding binary/non-monetary ones is incorrect. 
The counter-intuitive nature of this finding shows that it is very important to further develop an appropriate information-theoretic formalism, based on maximum-entropy statistical ensembles, aimed at identifying the key structural properties of economic networks.

Curiously, the road taken by Jan Tinbergen appears to lead to `physical' laws that are quite different from the ones originally postulated, and more similar to quantum statistics than gravitation. 
However, the deep epistemological reasons underlying Tinbergen's idea of introducing the Gravity Model of trade appear to be very appropriate, and persist throughout the more recent approaches.
This is, we believe, the most important legacy that Jan Tinbergen left us for the modern understanding of economic networks.

\begin{acknowledgement}
D. G. acknowledges support from the Dutch Econophysics Foundation (Stichting Econophysics, Leiden, the Netherlands) with funds from beneficiaries of Duyfken Trading Knowledge BV, Amsterdam, the Netherlands. 
\end{acknowledgement}

\end{document}